\documentclass[twocolumn,
amsmath,amssymb]{revtex4}
\usepackage{bm}
\usepackage[colorlinks=true,linkcolor=blue]{hyperref}
\usepackage{amsmath}
\usepackage{amssymb}
\usepackage{amsthm}
\usepackage{amsfonts}
\usepackage{enumerate}
\usepackage{latexsym}
\usepackage{ifpdf}
\usepackage{graphicx}
\expandafter\ifx\csname package@font\endcsname\relax\else
 \expandafter\expandafter
 \expandafter\usepackage
 \expandafter\expandafter
 \expandafter{\csname package@font\endcsname}%
 \fi
\begin{document}
\title{Strain dependent transport properties of the quasi two-dimensional correlated metal, LaNiO$_{3}$}
\author{E.J. Moon$^1$, B.A. Gray$^1$, M. Kareev$^1$, J. Liu$^1$, S.G. Altendorf$^2$, F. Strigari$^2$, L.H. Tjeng$^{2,3}$, J.W. Freeland$^4$ and J. Chakhalian$^1$}
\affiliation{$^1$Physics Department, University of Arkansas, Fayetteville, AR 72701, USA}
\affiliation{$^2$Physikalisches Institut, Universit\"{a}t zu K\"{o}ln, Z\"{u}lpicher Str., 50937 K\"{o}ln, Germany}
\affiliation{$^3$Max Planck Institute for Chemical Physics of Solids, N\"{o}thnitzerstr, 01187 Dresden, Germany}
\affiliation{$^4$Advanced Photon Source, Argonne National Laboratory, Argonne, IL 60439, USA}
\affiliation{$^5$Physics Department, University of Arkansas, Fayetteville, AR 72701, USA}

\begin{abstract}
We explore the electrical transport and magneto-conductance in quasi two-dimensional strongly correlated ultrathin films of LaNiO$_{3}$ (LNO) to investigate the effect of hetero-epitaxial strain on electron-electron and electron-lattice interactions from the low to intermediate temperature range (2K$\sim$170K). The fully epitaxial 10 unit cell thick films spanning tensile strain up to $\sim4\%$ are used to investigate effects of enhanced carrier localization driven by a combination of weak localization and electron-electron interactions at low temperatures. The magneto-conductance data shows the importance of the increased contribution of weak localization to low temperature quantum corrections. The obtained results demonstrate that with increasing tensile strain and reduced temperature the quantum confined LNO system gradually evolves from the Mott into the Mott-Anderson regime.

\end{abstract}

\maketitle

Low dimensional strongly correlated electron systems composed of artificial layers of transition metal oxides (TMO) have attracted a great amount of attention and been an active area of research for many years due to their intriguing electronic and magnetic properties \cite{Tokura:00, Dagotto:01, Dagotto:05}. Despite extensive research efforts, limited experimental information is available about the behavior of correlated carries in reduced dimensions and their response to the magnitude and sign of a strain state. For example, in TMO thin films, since strain due to the lattice mismatch between the epitaxial film and substrate can modify the electronic bandwidth, the balance between the Coulomb repulsion energy $U$ and electron hoping $t$ can be used to control the metal-to-insulator transition (MIT). Rare-earth nickelates (RENiO$_{3}$, RE=La, Pr, Nd, ..., Lu) represent an ideal candidate systems, where the one-electron bandwidth derived from the Ni-O chemical bond length and the Ni-O-Ni bond angle primarily controls the electronic ground state \cite{Catalan:08}. In the ultrathin limit, this is further amplified by the intrinsic tendency of correlated metals to localize in 2D\cite{Imada:98}. Recently, extensive transport measurements have been reported on films of LaNiO$_{3}$ (LNO) to address the role of reduced dimensionality and externally applied electric fields on conductivity \cite{Scherwitzl:09, Son1:10}. A few studies have focused on the effects of disorder in strongly correlated electronic systems, such as the resistivity minima at low temperature $T_{\textnormal{\tiny{min}}}$ which reveals the presence of quantum corrections to the conductivity. This quantum effect has been reported for SrRuO$_{3}$ (SRO). SRO, another important candidate for the electrical conductivity of the metallic perovskites, has a comparable Ioffle-Regel limit ($k_{B}l\sim1$, $k_{B}$ is the Boltmann constant, and $l$ is the mean free path) to LNO \cite{Rajeev:91}. The upturn in the electronic resistivity of ferromagnetic SRO is driven by renormalization of electron-electron interactions (REEI) due to the strong internal field \cite{Herranz:03}, whereas in LNO weak localization (WL) is significant \cite{Son1:10, Herranz:04}. For LNO ultrathin films (few nanometers), limited systematic information on the contributions of REEI and WL in terms of strain is available.      
\begin{table*}[]
\renewcommand{\arraystretch}{1.2}
\begin{center}
\caption{Results of fits to the Eq. 1 over the temperature range from 2K to 170K for LNO films with thickness 10 u.c. grown on various substrates. The substrate parameters compared with LNO are also displayed.}
\label{aggiungi}
\centering %
\begin{ruledtabular}
\tiny{\begin{tabular}{c c c c c c c c }
Substrate& Lattice  &  Strain  & $\sigma_0$  & $a_1$ & $a_2$ & b & $\alpha$ \\
&Parameter (\AA)&with LNO ($\%$)& ($\Omega^{-1}$cm$^{-1}$) & ($\Omega^{-1}$cm$^{-1}$K$^{-1}$) & ($\Omega^{-1}$cm$^{-1}$K$^{-1/2}$) & ($\Omega$cmK$^{-\alpha}$)&\\
\hline
LSAT&3.87&0.78&2795.8&6.6553&76.113&2.1135$\times$10$^{-7}$&1.3099\\
STO&3.901&1.915&3050.4&4.9923&68.258&1.8915$\times$10$^{-7}$&1.3199\\
TSO&3.954&3.224&3729.5&6.3772&62.489&3.6222$\times$10$^{-7}$&1.2558\\
GSO&3.969&3.612&4509.8&10.065&30.387&2.6661$\times$10$^{-7}$&1.2915\\
\end{tabular}}
\end{ruledtabular}
\end{center}
\end{table*}

In this paper, we present results on the systematic investigation of d.c. and magneto-transport properties of 10 unit cell thick ($\sim$3.83 nm) high-quality epitaxial LNO films as a function of tensile strain. Unlike the bulk, at low temperatures the d.c. transport measurements reveal that depending on the magnitude of the strain quantum corrections to conductivity arise from two distinct sources: a dominant electron-electron interaction and weak localization. Additionally, magneto-conductance as a function of magnetic field corroborates the dominant role of WL induced from electron scattering at the reduced dimensional regime. Therefore, here we invesitgate the importance of epitaxial strain and its sign on metaliclity as the material reaches the quasi-2D limit in LNO.

\begin{figure}[b]
  \centering
  \includegraphics[width=\columnwidth]{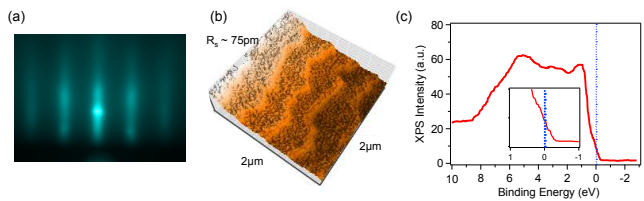}
  \caption{\label{fig:fig1}(Color online) (\emph{a}) The HP-RHEED image of LNO/STO after growth. (\emph{b}) The AFM micrograph (2$\times$2 $\mu$m$^{2}$) of the 10 u.c. film of LNO/STO. Its rms surface roughness of the deposited layer is $\sim$0.75 pm. (\emph{c}) XPS spectrum of a 10 u.c. LNO/STO. The region nearby Fermi-level is zoomed in the inset.} 
\end{figure} 

Epitaxial LNO ultra-thin films on (LaAlO$_3$)$_{0.3}$(Sr$_2$AlTaO$_6$)$_{0.7}$ (LSAT), SrTiO$_3$ (STO), TbScO$_3$ (TSO), and GdScO$_3$ (GSO) substrates were grown via pulsed laser deposition (PLD) with in situ monitoring by a recently developed high pressure Reflection High-Energy Electron Diffraction (HP-RHEED) system, which operates in an O$_{2}$ background pressure of up to 400 mTorr. The HP-RHEED image of LNO/STO is shown in Figure 1 (a). The details regarding the 2-D layer by layer growth PLD for LNO films are given elsewhere \cite{Kareev}. After the completion of deposition, the films were annealed in one atmosphere of ultra-pure oxygen to minimize possible oxygen deficiency that adversely affect the conductivity \cite{Greene:98, Zhu:06}. AFM imaging revealed smooth surface morphology with a surface roughness of 80 pm or better (see Figure 1 (b)). The photoemission spectra were recorded at room temperature in a spectrometer equipped with a VG-Scienta R3000 electron energy analyzer and a Vacuum Generators twin crystal monochromatized Al-Ka ($h\nu$ = 1486.6 eV) source. As shown in Figure 1 (c), the band crossing of the Fermi-level is clearly observed, which affirms the metallicity of 10 unit cell LNO films. The XPS results are in good agreement with an earlier study that showed the Fermi level of LNO lies in the conduction band due to strong overlapping of O 2$p$ and Ni 3$d$ bands\cite{Greene:98, Lisauskas:08}. 

We begin with a discussion of the temperature-dependent d.c. transport of LNO ultra-thin films. At finite temperatures, the resistivity as a function of temperature can be conventionally defined as $\rho(T)$=$\rho_{0}$+$AT^{\alpha}$, where $\rho_{0}$ is the temperature-independent residual resistivity. $A$ is a constant, and the power exponent $\alpha$ depends on the details of the scattering mechanism. Within the framework of Fermi-liquid theory, the Coulomb interaction yields $\alpha$ = 2 (or $T^2-$ dependence) \cite{Allen:92, Rivadulla:03}. However in complex oxides at lower temperatures, the carriers can localize, and a metal-insulator transition may occur in these marginal metals. The nature of the transition is then conventionally explained by the emergence of quantum corrections to the conductivity (QCC). The quantum corrections are derived from two important mechanisms: weak localization (WL), a self-interference effect, and electron-electron interactions (EEI)\cite{Ghantmakher}. 

\begin{figure}[b]
  \centering
  \includegraphics[width=\columnwidth]{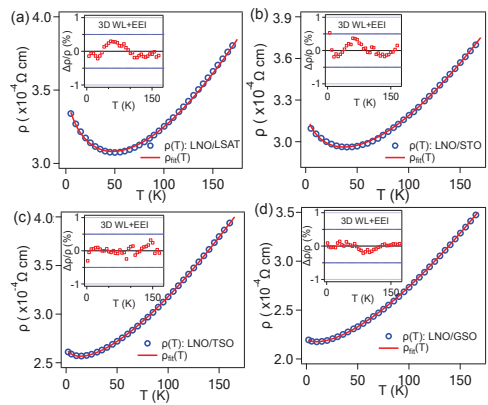}
  \caption{(Color online) Resistivity as a function of temperature for a LNO 10 u.c. film grown on (\emph{a}) a LSAT substrate, (\emph{b}) a STO substrate, (\emph{c}) a TSO substrate, and (\emph{b}) a GSO substrate with each fitting curve. Insets: Corresponding temperature dependence of the deviation from the fittings in the 3D limit.} 
\end{figure} 

Figure 2 shows the resistivity of LNO films on the four different substrates along with the fitting curves. The temperature dependent resistivity was
measured from 300K to 2K in the $van$ $der$ $Pauw$ geometry with a commercial physical properties measurement system (PPMS, Quantum Design). The resistivity upturn is clearly visible at low temperatures. The upturn can be described by considering both the EEI and WL contributions. Based on the localization-interaction model for a disordered metallic (e.g. `marginal' metal) system in the 3D limit, the temperature dependence of resistivity is given by \cite{Lee:85, Reghu:94, Eguchi:09}
\begin{equation}
\rho(T) = \frac{1}{\sigma_0 + a_{1}T^{p/2}+ a_{2}T^{1/2}} + bT^{\alpha},
\end{equation}
where $\sigma_{0}$ implies the classical temperature independent Drude conductivity, $a_{1}$ takes into account the 3-D WL contribution, and the last term, $a_{2}$, introduces the EEI in transport. The variable $p$ in the second term is an exponent which describes localization effects. It is well known that $p=2$ implies that the dominance of electron-electron interactions, while $p = 3$ is attributed to the electron-phonon scattering\cite{Lee:85}. As seen in Figure 2, for each LNO film, the value of $p$=2(0.003) results from the fitting to Eq. 1. Given the ultra-thin nature of the epitaxial LNO films, an attempt to fit the experimental data in the 2-D limit (i.e. including the $\ln{T}$ term instead of $a_{1}T^{p/2}+a_{2}T^{1/2}$ in Eq. 1)\cite{Lee:85} was also made, but the worse $\chi^{2}$ (1$\sim$2 order of magnitude of difference) confirmed that the 3-D fit is more appropriate. The magneto-conductance data, discussed later in the paper, corroborates the 3-D nature of the electronic transport. Thus, the best fit to the data for the 10 u.c. LNO films testifies to the 3-D nature of the localization. 

The temperature range for fitting to Eq. 1 was extended up to intermediate temperatures ($\sim$170K), because the upturn minima $T_{\textnormal{\tiny{min}}}$ of the four films vary between $\sim$7K and $\sim$50K. To investigate the effect of quantum corrections over the same temperature range and consider the same factors for all four films, the fit range includes  both the rise in the resistivity at low temperature and the intermediate metallic phase. In addition, to demonstrate the validity of the fitting parameters, the deviation plot defined as $\Delta\rho/\rho$ = ($(\rho_{\textnormal{\tiny{obs}}}(T)-\rho_{\textnormal{\tiny{fit}}}(T)$)/$\rho_{\textnormal{\tiny{obs}}}(T)$)$\times$100 is shown in Figure 2 (a)-(d), and the fitting parameters for the different substrates that correspond to varying the level of the epitaxial biaxial tensile strain on LNO are displayed in Table 1. 
\begin{figure}[t]
  \centering
  \includegraphics[width=\columnwidth]{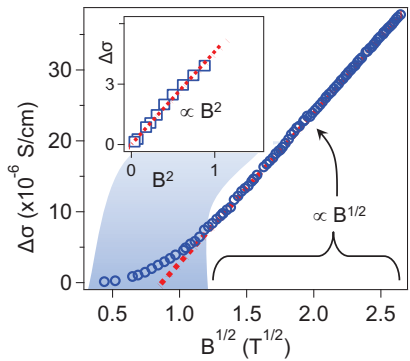}
  \caption{(Color online) Magnetoconductivity varies in field with its different power factor at the same system. The dotted lines are fits, to $B^2$ in the low field, and to $B^{1/2}$ in the high field regime.} 
\end{figure}

\begin{figure*}[t]
  \centering
  \includegraphics[width=12cm]{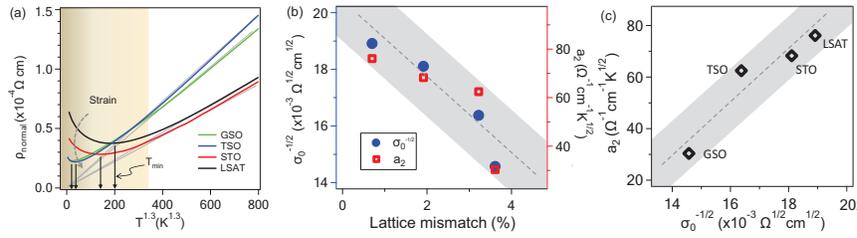}
  \caption{(Color online) (\emph{a}) $\rho_{\textnormal{\tiny{normal}}}(T)$ versus $T^{\alpha}$ ($\alpha$=1.3 or 4/3 as shown in Table 1 in the range of 2K$\sim$170K) of LaNiO$_{3}$ 10 u.c. films grown on different substrates to introduce various strain mismatches. The curvy arrows (gray color) across the four curves of resistivity show an evolution of the shape of upturn as a function of strain. The black-colored arrows under the curves indicate the temperature in which the resistivity is minimum (the position of the upturn). (\emph{b}) The coefficient $a_2$ versus the $\sigma_{0}^{-1/2}$. (\emph{c}) $a_{2}$ and $\sigma_{0}^{-1/2}$ as a function of strain mismatch between a LNO film and a substrate (see Table 1).} 
\end{figure*}
Information on the nature of localization regardless of quantum corrections can be obtained from magneto-conductance (MC) measurements \cite{Lee:85, Reghu:94, Ahlskog:96}. Specifically for a 3D system with dominant WL, the field dependent correction to the conductivity $\Delta\sigma(H,T)$ takes on a simple power law in limiting fields\cite{Lee:85, Bergmann:84}. For small applied fields, the MC is $\propto B^2$ ($g\mu_{B}B$$\textless$$k_{B}T$), and at the limit of high magnetic field the MC is $\propto B^{1/2}$ ($g\mu_{B}B$$\textgreater$$k_{B}T$) \cite{Reghu:94}. We performed MC measurements on the LNO sample grown on the GSO substrate (+4\% tensile strain) in an applied magnetic field of up to 7T parallel to a substrate \cite{footnote}. As seen in Figure 3, the magneto-conductance data taken at 2K follows the anticipated power law behavior at both limits. The upper limit for the low field dependence corresponds to $B = 1.3$T at 2K. The MC measurement provides additional information about the nature how the low $T$ localization largely attributed to weak localization, which is modulated by the magnitude of strain. Again, the observation of this field dependent MC supports the presence of weak localization of charge carriers in metallic films. The strain alters the distortion of the NiO$_{6}$ octahedra, and varies the $d$-band transport along with other factors that may contribute to interaction mechanisms. 

We now focus on the second power exponent term, $\sim$ $T^{\alpha}$, of Eq. 1 to obtain the contribution to the low $T$ quantum corrections from the metallic phase up to the intermediate temperature range. Figure 4 (a) shows the $\rho_{\textnormal{\tiny{normal}}}$ temperature dependence for LNO under increasing tensile strain from 2K to 170K; here the value of $\rho_{\textnormal{\tiny{normal}}}$ is defined as $\rho_{\textnormal{\tiny{normal}}}$=$\rho(T)-\rho_{\textnormal{\tiny{con}}}$ = $\rho_{\textnormal{\tiny{QCC}}}$+$T^{\alpha}$, where $\rho_{\textnormal{\tiny{con}}}$ is a constant to rescale each resistivity curve to cross zero resistivity at 0K. As seen in Figure 2 and Table 1, the power $\alpha$ yields 1.29(0.03) ($\sim$ 4/3). 

Upon approaching the low temperature range, the d$T^{4/3}$/d$\rho_{\textnormal{\tiny{normal}}}(T)$ gradually changes with strain. As shown in Figure 4, the region within the yellow-filled enclosed box exhibites significant changes of the d$T^{4/3}$/d$\rho_{\textnormal{\tiny{normal}}}(T)$ in terms of strain. Furthermore, the minimum of resistivity $T_{\textnormal{\tiny{min}}}$ decreases with strain as illustrated by the black arrows under the resistivity curves in Figure 4 (a). This resistivity upturn (or MIT) may result from several plausible causes. For instance, it is well known that after reaching the Ioffe-Regel limit, in which the mean free path equals the interatomic spacing the film may undergo a transition from the metallic to insulating ground state\cite{Ioffe}. The thickness of the LNO sample (few nanometers around the critical thickness) will also enhance the propensity towards localized behavior \cite{Son1:10}. Here, we note however, that the minimum conductivity from the Mott-Ioffe-Regel limit (MIRL) \cite{Mott} of bulk LNO is $\sim$ 300 S/cm \cite{Son2:10}, whereas the conductivity of each strained LNO ultra-thin film from LSAT to GSO (upto the tensile strain of 4$\%$) is well above this MIRL. Given these results, the localization observed in the quality 10 u.c. films is not primarily driven by the intrinsic disorder and/or the reduced layer thickness (size effect). Instead, based on the strong correlation between the low $T$ localization and the magnitude of strain, the lattice mismatch which modifies the one-electron bandwidth is the prime source for such response. To further quantify this, we consider the relative contribution of the fitting parameters of Eq. 1 ($a_{1}$, $a_{2}$, and $\sigma_{0}$ (see Table 1)) vs. strain by means of the empirically derived relation reported in Ref. \cite{Herranz:04, Abrikosov:88}. As shown in Figure 4 (b)-(c), the parameter $a_{2}$ decreases with increasing tensile strain, indicating that EEI is suppressed with tensile strain, while the WL contribution ($a_{1}$ parameter) is clearly enhanced with strain. The relative contribution of localization and Coulomb interactions in the transport properties of the film might be controlled by strain. These trends of the fitting and scalling parameters may indicate that LNO ultrathin films are closer to Mott-Anderson materials, which can be determined by soft Hubbard gap \cite{Shinaoka}, than Mott materials. The synergetic contribution of these two effects is perhaps responsible for the peculiar low temperature dynamics with strain.   

We have found the quantum corrections to the conductivity for charge carriers in LaNiO$_{3}$ ultra-thin films and modeled the temperature dependent resistance in terms of three dimensional weak localization and electron-electron interactions in a disordered metallic system. The synergetic contribution of these two effects is perhaps responsible for the peculiar low temperature dynamics with strain. 

The authors thank Bogdan Dabrowski and Derek J. Meyers for useful discussions. JC was supported by DOD-ARO under the Grant No. 0402-17291 and NSF Grant No. DMR-0747808.


\begin{thebibliography}{999}
\bibitem{footnote} Based on the ratio of $a_{1}$ and $a_{2}$ as a function of strain, it is plausible to conclude that the LNO ultrathin film grown on GSO substrate is the best candidate for WL.
\end{thebibliography}

\end{document}